\documentclass[
prd
,showpacs ,amssymb,superscriptaddress,aps
]{revtex4}
\usepackage{graphicx}
\input{epsf}

\usepackage{amsmath,amssymb}
\usepackage{bm}
\usepackage{times}

\newcommand{\dalm}{\kern1pt\vbox{\hrule height 0.9pt\hbox{\vrule width 0.9pt
\hskip 2.5pt\vbox{\vskip 5.5pt}\hskip 3pt\vrule width 0.3pt}\hrule height 0.3pt}
\kern1pt}

\begin{document}



\title{Empirical formula of crustal torsional oscillations}

\author{Hajime Sotani}
\email{hajime.sotani@nao.ac.jp}
\affiliation{Division of Theoretical Astronomy, National Astronomical Observatory of Japan, 2-21-1 Osawa, Mitaka, Tokyo 181-8588, Japan}


\date{\today}

\begin{abstract}
Crustal torsional oscillations depend on not only crust properties but also the stellar mass and radius. Thus, one could extract  stellar information by identifying the observed frequencies of stellar oscillations with the crustal torsional oscillations. Owing to the confinement of torsional oscillations inside the crust region of neutron stars, we successfully derive an empirical formula for the fundamental crustal torsional oscillations as a function of the stellar mass, radius, the so-called slope parameter of the nuclear symmetry energy, and the angular index of oscillations, with which one can estimate the frequencies with high accuracy. This empirical formula could be valuable in both the astrophysics and nuclear physics communities. 
\end{abstract}

%
\maketitle
\section{Introduction}
\label{sec:I}

Neutron stars are one of the very few suitable environments that realize the extreme conditions where the density inside the star easily becomes greater than the nuclear saturation density, the magnetic fields inside and around the star can be quite strong, and the gravitational field becomes much stronger than that in the Solar System. So, via observables associated with neutron stars, one can expect to get crucial information that is quite difficult to obtain on Earth. In such attempts, the stellar mass and/or radius must be important information. In fact, the discovery of neutron stars with mass $\sim 2M_\odot$ has excluded some soft equations of state (EOSs) with which the expected maximum mass cannot reach $2M_\odot$ \cite{2Ma,2Mb}. Meanwhile, the stellar mass and radius may be weak information if one focuses on the crust properties, because the crust thickness is at most $\sim 10\%$ of the stellar radius, and the stellar mass and radius are not very sensitive to the crustal properties. In this case, the oscillation spectra of neutron stars could provide additional/alternative observable information to extract the interior properties of neutron stars. This technique is known as asteroseismology, which is similar to seismology for  the Earth and helioseismology for the Sun. So far, there have been many attempts to measure stellar properties via oscillation spectra, such as the stellar mass, radius, and EOS (e.g., Refs. \cite{AK1996,STM2001,SH2003,Doneva2013}), the imprint of quark matter inside the star \cite{SYMT2011}, and magnetic properties (e.g., Refs. \cite{Gabler2011,PA2012,Gabler2012a}). In addition to theoretical studies, there have been recent attempts to actually extract the stellar properties by fitting observational data (e.g., Refs. \cite{I2005,SW2005,SW2006}).

The structure of neutron stars strongly depends on the EOS for neutron star matter. Although the EOS (even for a high-density region) should be somehow constrained by terrestrial nuclear experiments, it is quite difficult to make a constraint on the EOS for regions with densities much higher than the saturation density, due to the nature of the saturation property of nuclear matter. Thus, there are still many uncertainties in the EOS for neutron star matter, especially for high-density region, which leads to the difficulty of determining the exact structure of neutron stars. Even in such a situation, we have a consensus about the  conceptual structure of neutron stars \cite{NS}. A thin ocean may exist under the atmosphere at the outermost part of neutrons stars. Then, matter inside the ocean forms a Coulomb lattice and behaves as a solid, which corresponds to the crust region of neutron stars. Also, the neutrons confined in the nuclei begin to drip out when the density becomes greater than $4\times 10^{11}$ g/cm$^3$. A crust region with a density lower (higher) than this critical density corresponds to the outer (inner) crust. It is thought that the majority of the crust region is composed of spherical nuclei forming a body-centered cubic lattice, but it may be possible to deform the shape of nuclei at the bottom of the crust region (such as the so-called pasta structure) \cite{LRP1993,O1993}. Anyway, the matter eventually becomes uniform when the density becomes more than the critical density (which depends on the EOS), which is around $(0.5-1)$ times the nuclear saturation density. The region inside the crust region becomes a fluid core, whose properties play an important role in determining the radius and mass of the neutron star.

In the case when one attempts to extract the information around the saturation density via the observations of a neutron star, the phenomena associated with the crust region become crucial. On the other hand, as mentioned later in Sec. \ref{sec:II}, the EOS around the saturation point can be expanded as a function of the baryon number density and neutron excess with the saturation parameters, such as in Eq. (\ref{eq:w}). In particular, since the incompressibility ($K_0$) and the so-called slope parameter of the nuclear symmetry energy ($L$) among the saturation parameters are relatively difficult to determine via  terrestrial nuclear experiments, astronomical observations should be helpful to constrain $K_0$ and $L$. In practice, the crustal oscillations in neutron stars could tell us the imprint of the crustal properties, which depend on $K_0$ and/or $L$. This is a motivation for considering the crustal torsional oscillations.

From the asteroseismological point of view, gravitational waves must be the most promising astronomical information, although their direct detection has not yet occurred. On the other hand, the discovery of the quasiperiodic oscillations in giant flares or in smaller bursts observed from soft-gamma repeaters \cite{I2005,SW2005,SW2006,WS2006,Hupp2014a,Hupp2014b} gives the  opportunity to adopt the asteroseismology to extract stellar information. In practice, in order to explain these quasiperiodic oscillations theoretically, there have been many attempts in terms of the crustal torsional oscillations \cite{I2005,SW2005,SW2006,SA2007,SW2009,Sotani2011,GNHL2011,Sotani2014,DSB2014} and/or magnetic oscillations of neutron stars \cite{Levin2006,Lee2007,SKS2007,SKS2008,Sotani2008b,Sotani2009,CSF2009,CK2011,PL2013,GCFMS2013,PL2014,Gabler2011,PA2012,Gabler2012a}. Furthermore, the possibility of ruling out a specific strange star model by using the observed quasiperiodic oscillations was also suggested \cite{WR2007}. Through such recent analyses about the oscillations of magnetized neutron stars, it has been found that the magnetic oscillations can be coupled with the crustal oscillations when the strength of magnetic fields is larger than a critical field strength, such as $\sim 10^{15}$ G \cite{CK2011,Gabler2011,Gabler2012a}. That is, one has to consider the magnetoelastic torsional oscillations for strongly magnetized neutron stars. In addition to the effect of magnetic fields, the superfluidity inside the star could also play an important role in neutron star oscillations \cite{PL2013,Gabler2013,PL2014}. Even so, since there are still many uncertainties in the magnetic distribution and strength and in the physics of superfluidity inside neutron stars, an analysis of pure crustal torsional oscillations could be valuable. In fact, identifying the observed quasiperiodic oscillations with the crustal torsional oscillations allows one to make a constraint on the saturation parameters in nuclear matter \cite{SW2009,DSB2014,SNIO2012,SNIO2013a,SNIO2013b,SIO2016}.

However, in general, the determination of the frequencies of crustal torsional oscillations is not so easy, because one should solve the eigenvalue problems in the relativistic framework. Additionally, the frequencies could depend on not only the stellar mass and radius but also the crust EOS. That is, if an empirical formula for the frequency of crustal torsional oscillations does exist, it is useful to identify the observations with the crustal torsional oscillations without any complicated tasks. With respect to this demand, in this paper we successfully derive such an empirical formula --as a function of the stellar mass, radius, $L$, and the angular index of oscillations $\ell$-- that can estimate the fundamental frequencies with high accuracy. We adopt geometric units, $c=G=1$, where $c$ and $G$ denote the speed of light and the gravitational constant, respectively, and the metric signature is $(-,+,+,+)$ in this paper.

\section{Torsional oscillations}
\label{sec:II}

Torsional oscillations are axial-type oscillations, which do not involve density variations. Therefore, it is possible to estimate the frequencies with surgical precision even with the relativistic Cowling approximation, where the metric perturbations are neglected during the oscillations. The restoring force of torsional oscillations is the shear stress due to the  elasticity of neutron star matter. That is, the torsional oscillations are assumed to be excited only in the crust region of a neutron star, because the matter in the core region could behave as a fluid \cite{comment1}. This is an advantage in attempts to extract information of the  crust region via the oscillation spectrum. In fact, there are many uncertainties in the EOS for the core region, which have yet to  be constrained from terrestrial nuclear experiments. However, due to the independence of the torsional oscillations from the core EOS, one can directly discuss the relationship between the spectra of torsional oscillations and crust properties irrespective of any uncertainties in the core region.

The neutron star models are constructed by integrating the well-known Tolman-Oppenheimer-Volkoff (TOV) equation together with the appropriate EOS. In general, one obtains an equilibrium one-parameter family of neutron star models (such as the  central density or stellar mass) by integrating the TOV equation from the stellar center up to the stellar surface, if one adopts an EOS expressing not only the crust (low-density) region but also the core (high-density) region. In this case, by definition, one cannot avoid the uncertainties in the core region. Additionally, the incorporation of the uncertainties in the core region would be inevitable, when one considers the stellar oscillations associated with the core region and/or the stellar magnetic fields penetrating the core region. On the other hand, if one only has an interest in properties of the crust region such uncertainties can be avoided by integrating the TOV equation inward from the stellar surface to the crust basis \cite{IS1997}. We remark that in this scheme one has to prepare two parameters --i.e., the stellar mass ($M$) and radius ($R$)-- to construct the neutron star crust models, which might be a weak point in the scheme with inward integration. 

To construct the crust models, one needs to prepare the EOS for the crust region. In particular, we adopt the phenomenological EOS proposed by Oyamatsu and Iida \cite{OI2003,OI2007} as in the previous studies \cite{SNIO2012,SNIO2013a,SNIO2013b,SIO2016}. Hereafter, we refer to this phenomenological EOS as OI-EOS. With any EOSs, the bulk energy per nucleon can be expanded around the saturation point of symmetric nuclear matter at zero temperature as a function of baryon number density $n_{\rm b}$ and neutron excess $\alpha$ \cite{L1981}:
\begin{equation}
w = w_0  + \frac{K_0}{18n_0^2}(n_{\rm b}-n_0)^2 + \left[S_0 + \frac{L}{3n_0}(n_{\rm b}-n_0)\right]\alpha^2. \label{eq:w}
\end{equation}
In this expansion, $w_0$, $n_0$, and $K_0$ correspond to the saturation energy, saturation density, and incompressibility of the symmetric nuclear matter, while $S_0$ and $L$ are parameters associated with the nuclear symmetry energy, i.e., $S_0$ is the symmetry energy at $n_{\rm b}=n_0$ and $L$ is the so-called slope parameter of the symmetry energy. Among these five parameters, $w_0$, $n_0$, and $S_0$ are well constrained via terrestrial nuclear experiments due to the nature of the saturation property of nuclear matter. On the other hand, it is relatively more difficult to determine the remaining two parameters $K_0$ and $L$, because one has to obtain nuclear data in the wide range of densities around the saturation point. So, the OI-EOS is designed to reproduce Eq. (\ref{eq:w}) in the limit of $n_{\rm b}\to n_0$ and $\alpha\to 0$, where the values of $w_0$, $n_0$, and $S_0$ are optimized to recover the empirical nuclear data for stable nuclei for given values of $K_0$ and $L$. We remark that one could fairly commonly examine the crust properties even with the phenomenological EOS adopted here, because the density at the crust basis is only half of the saturation density up to at most the saturation density, whose properties can be described well by the saturation parameters.

At last, in order to construct the crust equilibrium models, we have to prepare four parameters, i.e., two parameters for EOSs ($K_0$ and $L$) and two parameters for neutron star models ($M$ and $R$). In particular, we consider the stellar models in the ranges of $1.4\le M/M_\odot \le 1.8$ and $10 \le R\le 14$ km. We remark that these mass and radius ranges are a little narrow, compared to the observed and theoretical values \cite{2Ma,2Mb,Glen2000}. Meanwhile, as in the previous studies \cite{SNIO2012,SNIO2013a,SNIO2013b,SIO2016}, we adopt the parameter ranges $0\le L \le 160$ MeV and $180\le K_0\le 360$ MeV, which not only well reproduce the experimental data for stable nuclei but also effectively cover even extreme cases \cite{OI2003}. In practice, in this paper we adopt the same EOS parameters as those shown in Table 1 in Ref. \cite{SNIO2013b}.

The elasticity (which is strongly associated with a restoring force of torsional oscillations) is characterized by the shear modulus $\mu$, which is principally determined by the lattice energy due to the Coulomb interaction. Since the nuclei in most parts of the crust region are generally considered to form a body-centered cubit lattice, in this paper we adopt the corresponding shear modulus, which was formulated in Ref. \cite{SHOII1991} with the ion number density ($n_i$), charge number of the ion ($Z$), and the radius of a Wigner-Seitz cell ($a$):
\begin{equation}
  \mu = 0.1194\,\frac{n_i (Ze)^2}{a}. \label{eq:mu0}
\end{equation}
It should be noted that this formula was derived in the limit of zero temperature of the shear modulus obtained from Monte Carlo calculations, where each nucleus is assumed to be a charged point particle \cite{OI1990}. Subsequently, the additional effects in the shear modulus are also taken into account, i.e.,  the contribution of the lattice phonons \cite{Baiko2011}, the modification due to the electron screening \cite{KP2013}, and the possibility of changes of the effective shear modulus by considering randomly oriented polycrystalline matter \cite{KP2015}. It seems that the modification due to the lattice phonons is negligible \cite{Baiko2011,KP2013}, while the effects of the electron screening and the randomly oriented polycrystalline matter can lead to $\sim 10\%$ and $\sim 30\%$ reductions of the shear modulus, respectively, which correspond to $\sim 5\%$ and $\sim 15\%$ reductions in torsional frequencies \cite{KP2013,KP2015}. In addition, here we consider the effective shear modulus for zero temperature, but the shear modulus given by Eq. (\ref{eq:mu0}) is a good approximation for neutron star  temperatures below $\sim 10^8$ K \cite{GNHL2011}. Nevertheless, in this paper, we adopt the canonical formula [Eq. (\ref{eq:mu0})] for the effective shear modulus.

The frequencies of the crustal torsional oscillations can be determined by solving the eigenvalue problem. That is, one should integrate the perturbation equation derived by linearizing the equation of motion together with the appropriate boundary conditions imposed at the crust basis and stellar surface \cite{ST1983,SKS2007}. The concrete perturbation equation and the boundary conditions can be seen in our previous studies \cite{SNIO2012,SNIO2013a,SNIO2013b,SIO2016}. Furthermore, we also incorporate the effect of neutron superfluidity on the torsional oscillations in the same way as in Refs. \cite{SNIO2013a,SNIO2013b}. In fact, it is considered that the neutrons confined in the nuclei star drip out when the rest-mass density becomes more than $\sim 4\times 10^{11}$ g/cm$^3$, and that a portion of the dripped neutrons may act as a  superfluid. In the calculations, we adopt the ratio of superfluid neutrons to the dripped neutrons based on the band calculations \cite{C2012}. According to the result in Ref. \cite{C2012}, such a ratio is in the range of $10-30\%$,  depending on the density.

\section{Deriving the empirical formula of crustal torsional oscillations}
\label{sec:III}

In the Newtonian case, it is known that the fundamental frequencies of crustal torsional oscillations with angular index $\ell$, which are denoted by ${}_\ell t_0$, are expressed as
\begin{equation}
  {}_\ell t_0 \approx \frac{2\pi v_s\sqrt{\ell(\ell+1)}}{R},   \label{eq:Newton}
\end{equation}
where $v_s$ denotes the typical shear velocity \cite{HC1980}. On the other hand, ${}_\ell t_0$ must depend on the crust EOS. In fact, via the relativistic perturbation approach mentioned in Sec. \ref{sec:II}, we have already shown that the fundamental frequencies for a given stellar mass and radius can be expressed as a function of $L$ almost independently of $K_0$ (see Fig. 3 in Ref. \cite{SNIO2013b}), i.e.,
\begin{equation}
  {}_\ell t_0 = c_{\ell, 0} - c_{\ell, 1}\left(\frac{L}{100\; {\rm MeV}}\right) + c_{\ell, 2} \left(\frac{L}{100\; {\rm MeV}}\right)^2,  \label{eq:lt0}
\end{equation}
where $c_{\ell,0}$, $c_{\ell,1}$, and $c_{\ell,2}$ are positive fitting constants that depend on the index $\ell$ and the stellar mass and radius \cite{SNIO2012}. We remark that this fitting can estimate the frequencies of the fundamental torsional oscillations within a few percent except for the unrealistic case with a very small value of $L$ \cite{SNIO2013b,SIO2016}. In the previous studies, we also found the trend that the frequencies become small for larger stellar radii and more massive neutron  star models. However, a practicable empirical formula for the frequencies of crustal torsional oscillations does not exist, in spite of several discussions about the relationship between the crust properties and frequencies of torsional oscillations. Thus, we try to derive the empirical formula of ${}_\ell t_0$ as a function of not only $L$ but also $\ell$, $M$, and $R$, which must be valuable in both the astrophysics and nuclear physics communities.

\begin{figure}
\begin{center}
\includegraphics[scale=0.5]{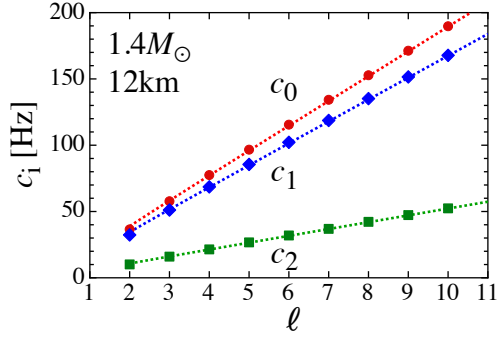} 
\end{center}
\caption{
The values of the coefficients in Eq. (\ref{eq:lt0}) with various angular indices $\ell$ for the neutron star model with $M=1.4M_\odot$ and $R=12$ km. The dotted lines correspond to the fitting lines with Eq. (\ref{eq:fit1}).
}
\label{fig:ci-M14R12}
\end{figure}

\begin{figure*}
\begin{center}
\begin{tabular}{ccc}
\includegraphics[scale=0.42]{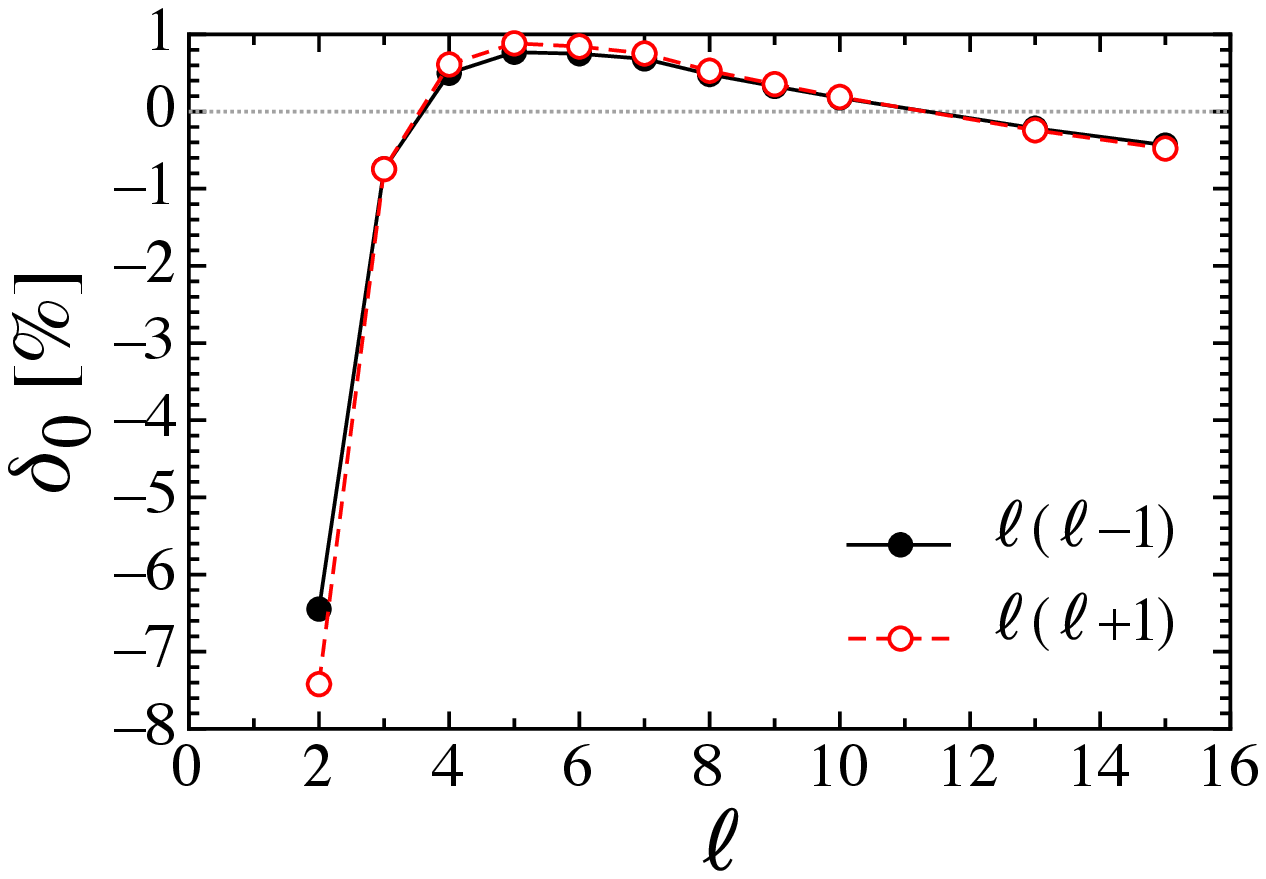} &
\includegraphics[scale=0.42]{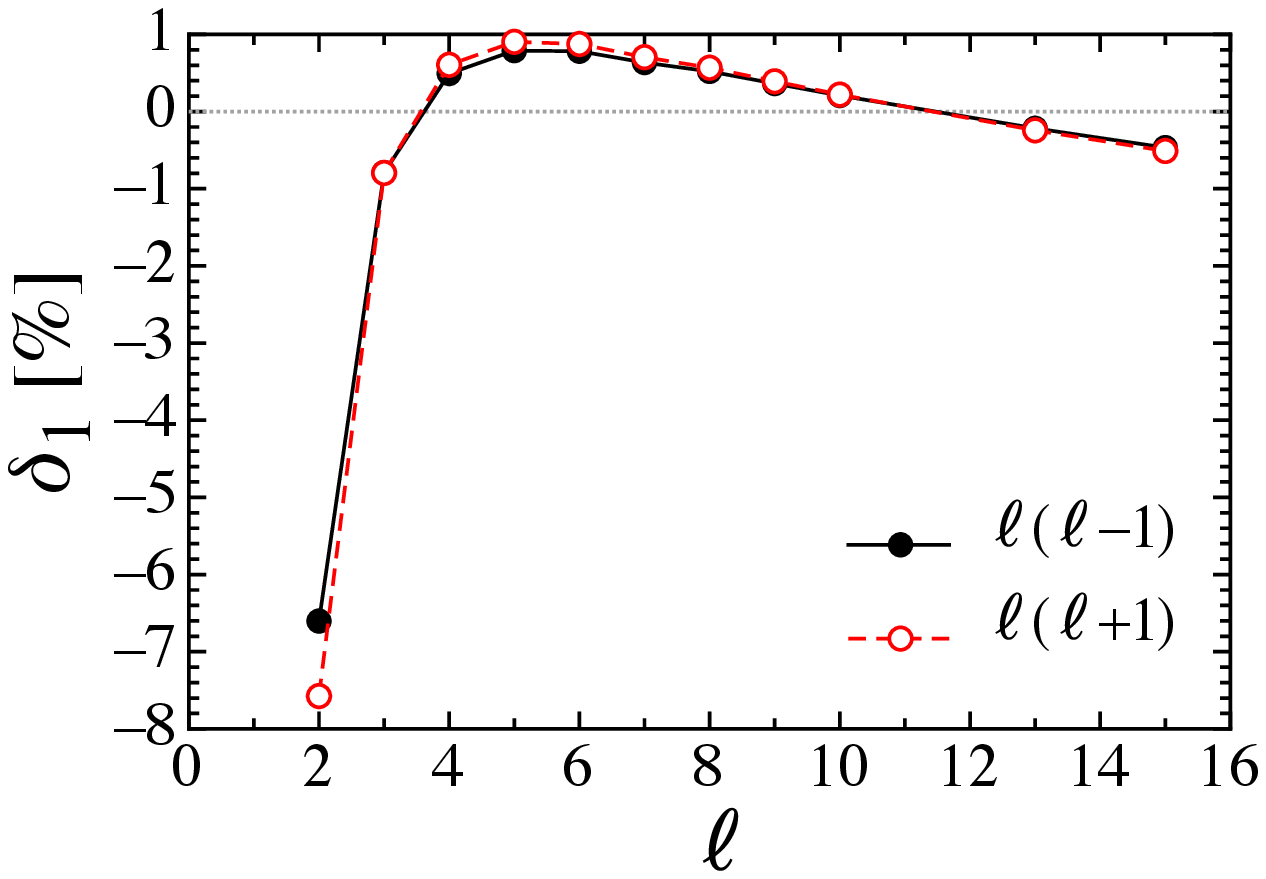} &
\includegraphics[scale=0.42]{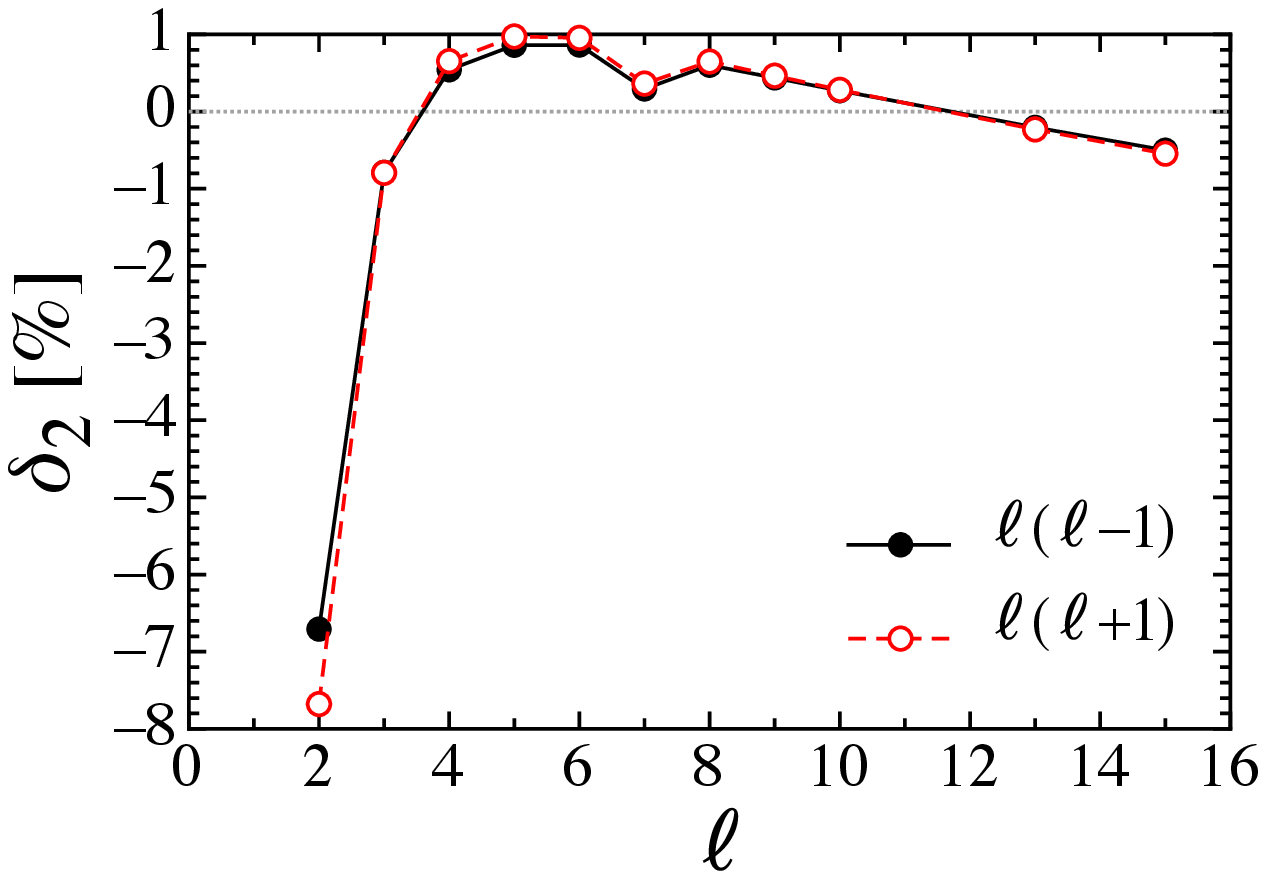} 
\end{tabular}
\end{center}
\caption{
Relative deviation of the coefficients in Eq. (\ref{eq:lt0}) from two different fitting formulas for the neutron star model with $M=1.4M_\odot$ and $R=12$ km. The filled and open circles correspond to the relative deviations calculated with Eqs. (\ref{eq:fit1}) and (\ref{eq:fit2}), respectively.  
}
\label{fig:delta}
\end{figure*}

First, we fit the dependence of $c_{\ell,i}$ on $\ell$. In Fig. \ref{fig:ci-M14R12}, we show the exact values of $c_{\ell,i}$ with various angular indices $\ell$ for the stellar model with $M=1.4M_\odot$ and $R=12$ km, as a typical neutron star model. The fitting constants $c_{\ell,i}$ appear to be linear functions of $\ell$. But, through a trial and error process, we find that the dependence of $c_{\ell,i}$ on $\ell$ could be expressed better with functional forms, such as
\begin{gather}
  c_{\ell,i} = d_{i0} + d_{i1} \sqrt{\ell(\ell-1)},  \label{eq:fit1} \\
  c_{\ell,i} = d_{i0} + d_{i1} \sqrt{\ell(\ell+1)},  \label{eq:fit2}
\end{gather}
where $d_{i0}$ and $d_{i1}$ are some constants for $i=0$, 1, and 2 that depend on the stellar mass and radius. We remark that, if we assume that $c_{\ell,i}$ is just a linear function of $\ell$, i.e., $c_{\ell,i} = d_{i0} + d_{i1} \ell$, we cannot express the dependence of $d_{ij}$ on the stellar mass and radius very well, as discussed later. Using the above functional forms, the relative deviation can be estimated by
\begin{equation}
  \delta_i = \frac{c_{\ell,i}-\bar{c}_{\ell,i}}{c_{\ell,i}}, \label{eq:delta}
\end{equation}
where $c_{\ell,i}$ and $\bar{c}_{\ell,i}$ are the exact values in Eq. (\ref{eq:lt0}) and the expected values from the fitting (\ref{eq:fit1}) or (\ref{eq:fit2}), respectively, for each stellar model. In Fig. \ref{fig:delta}, we show the relative deviation calculated by Eq. (\ref{eq:delta}) for the stellar model with $M=1.4M_\odot$ and $R=12$ km, where the filled and open circles denote the values estimated with the fittings (\ref{eq:fit1}) and (\ref{eq:fit2}). From this figure, one can observe that both fitting formulas  work well. In particular, we can estimate the values of $c_{\ell,i}$ except for the case of $\ell=2$ with the fitting (\ref{eq:fit1}) or (\ref{eq:fit2}) within an accuracy of less than $1\%$. Even so, careful observation shows that the fitting (\ref{eq:fit1}) is better than Eq. (\ref{eq:fit2}). Thus, in this paper we adopt the fitting (\ref{eq:fit1}) to express the dependence of $\ell$.

\begin{figure*}
\begin{center}
\begin{tabular}{ccc}
\includegraphics[scale=0.42]{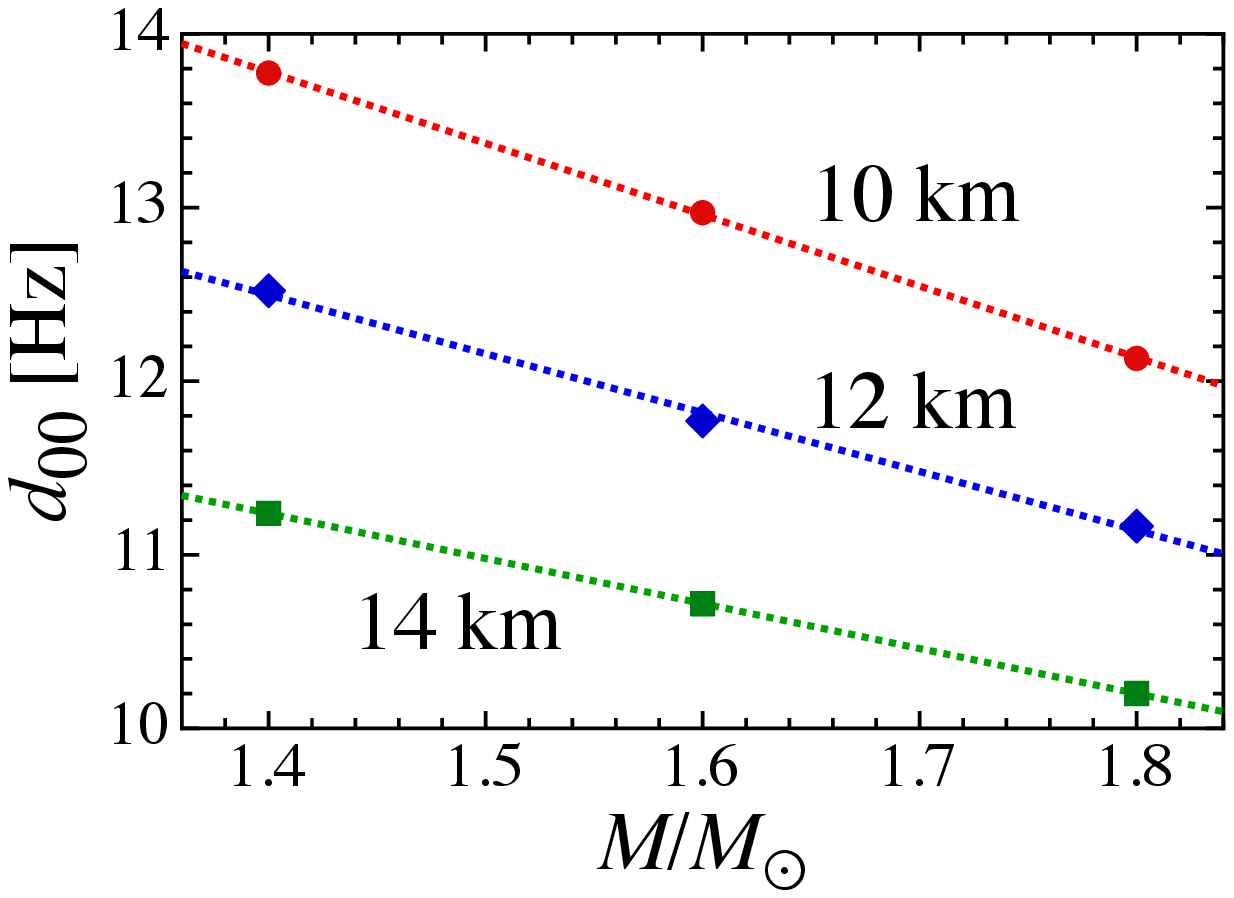} &
\includegraphics[scale=0.42]{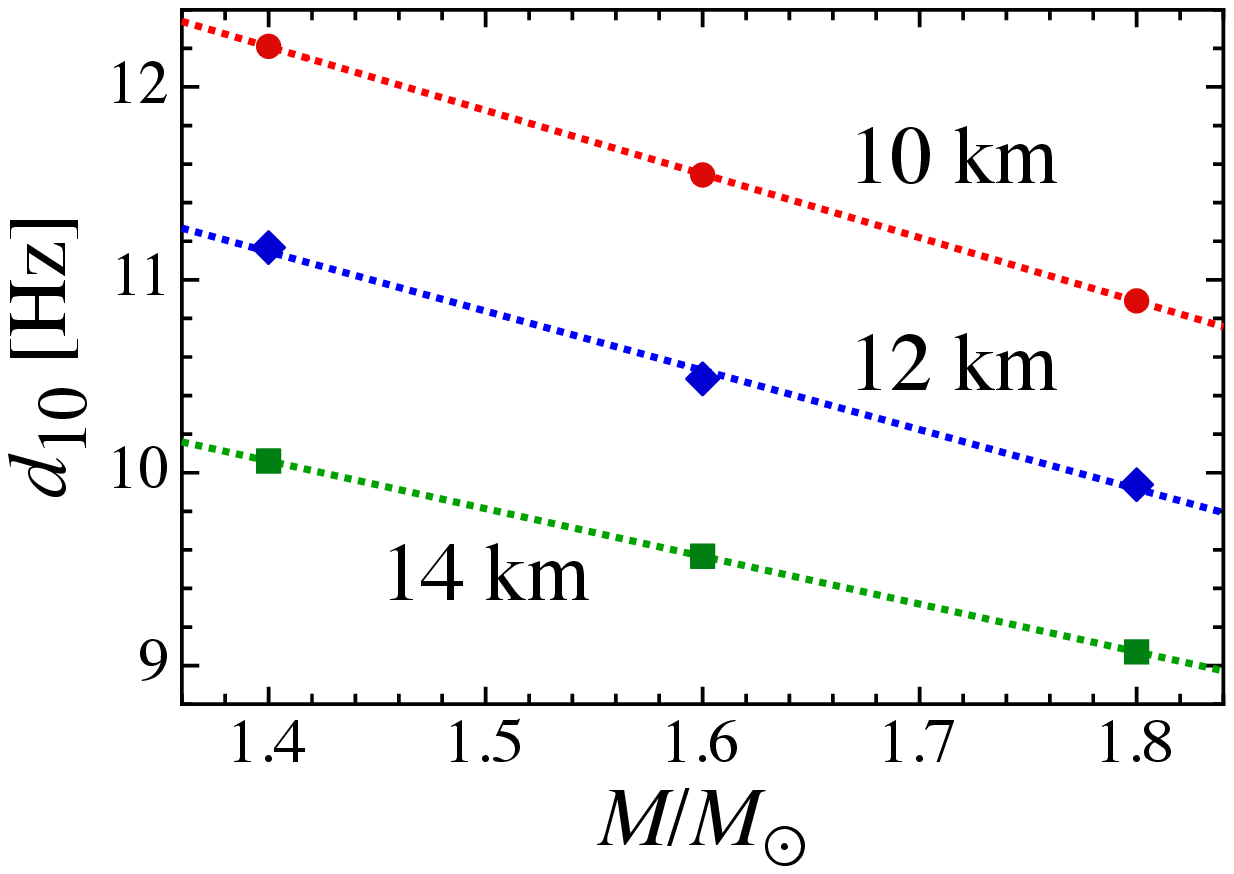} &
\includegraphics[scale=0.42]{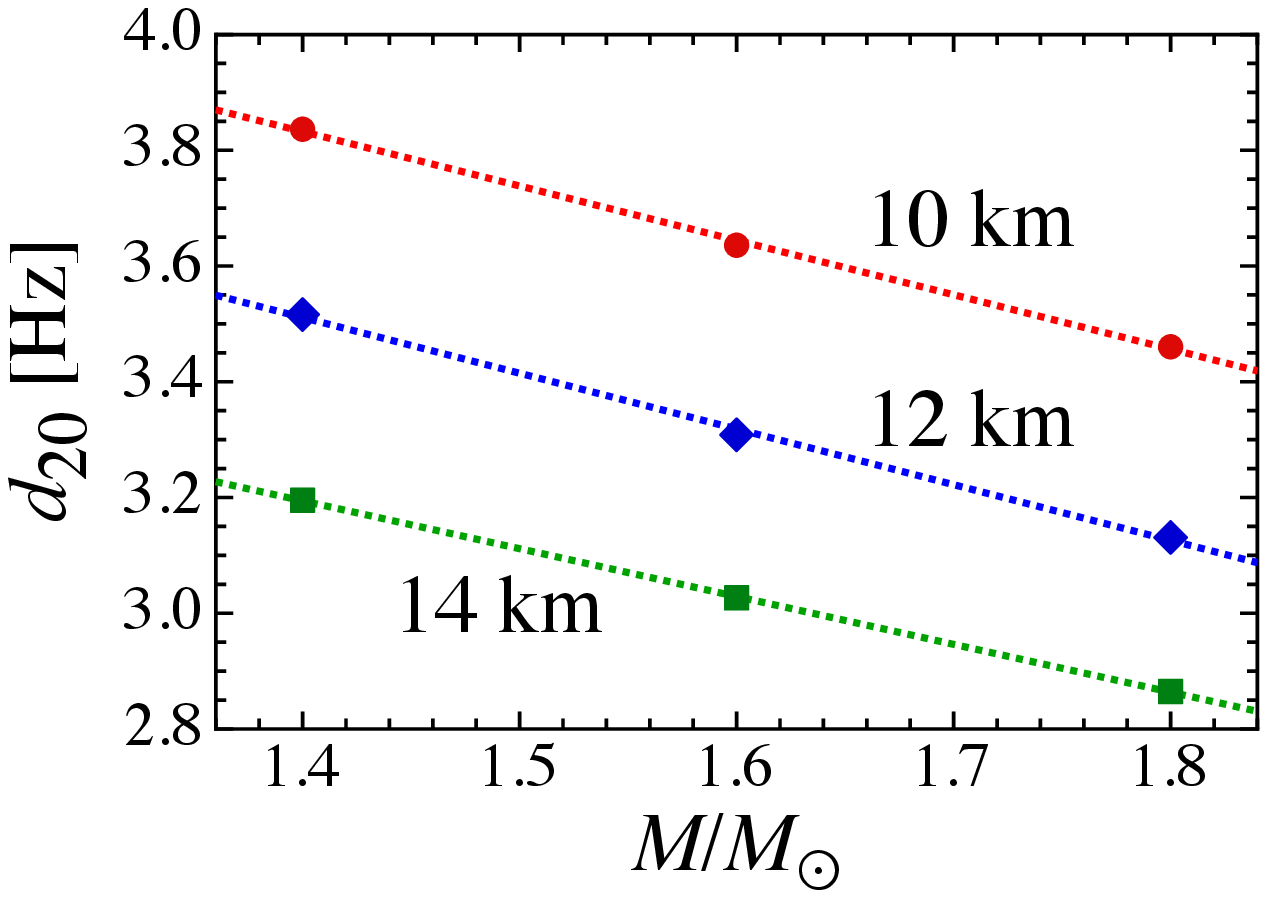} \\ 
\includegraphics[scale=0.42]{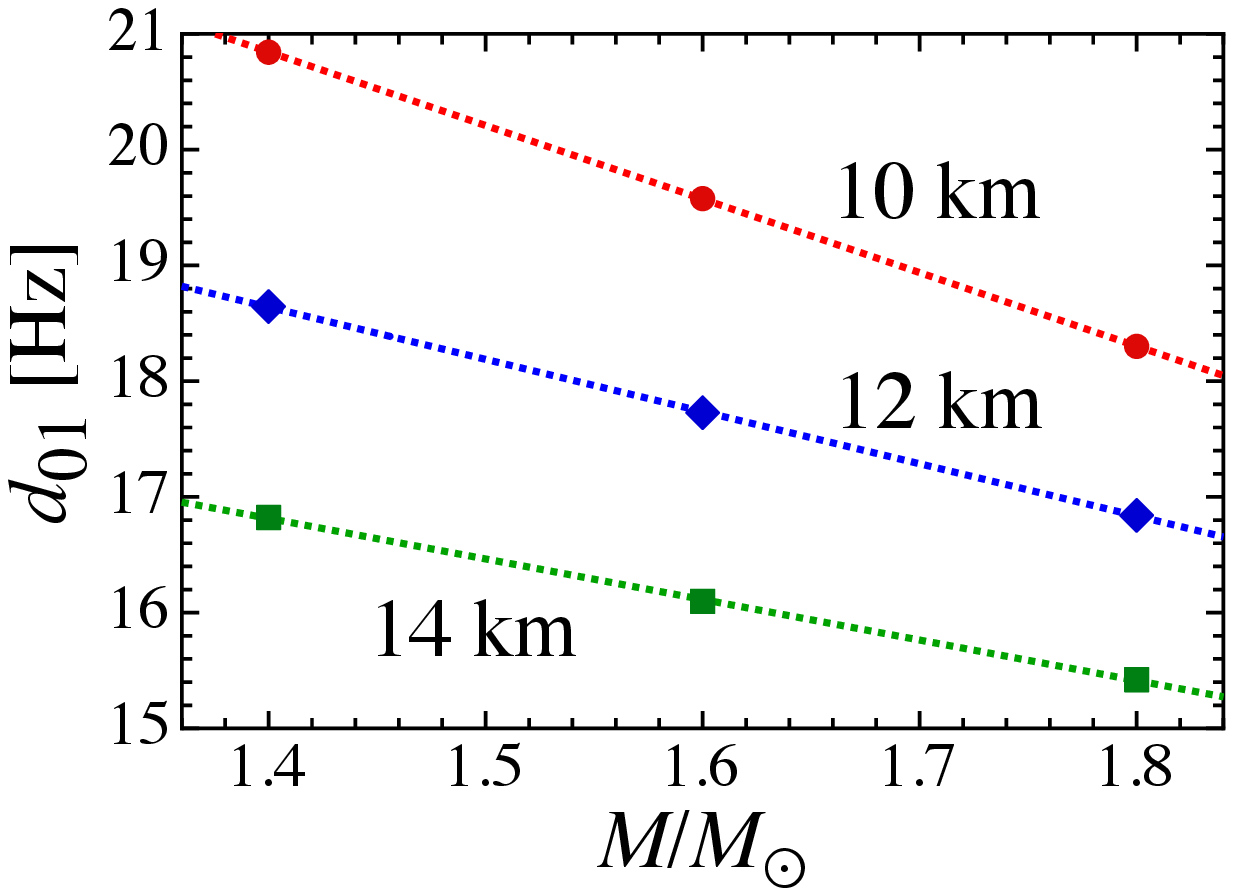} &
\includegraphics[scale=0.42]{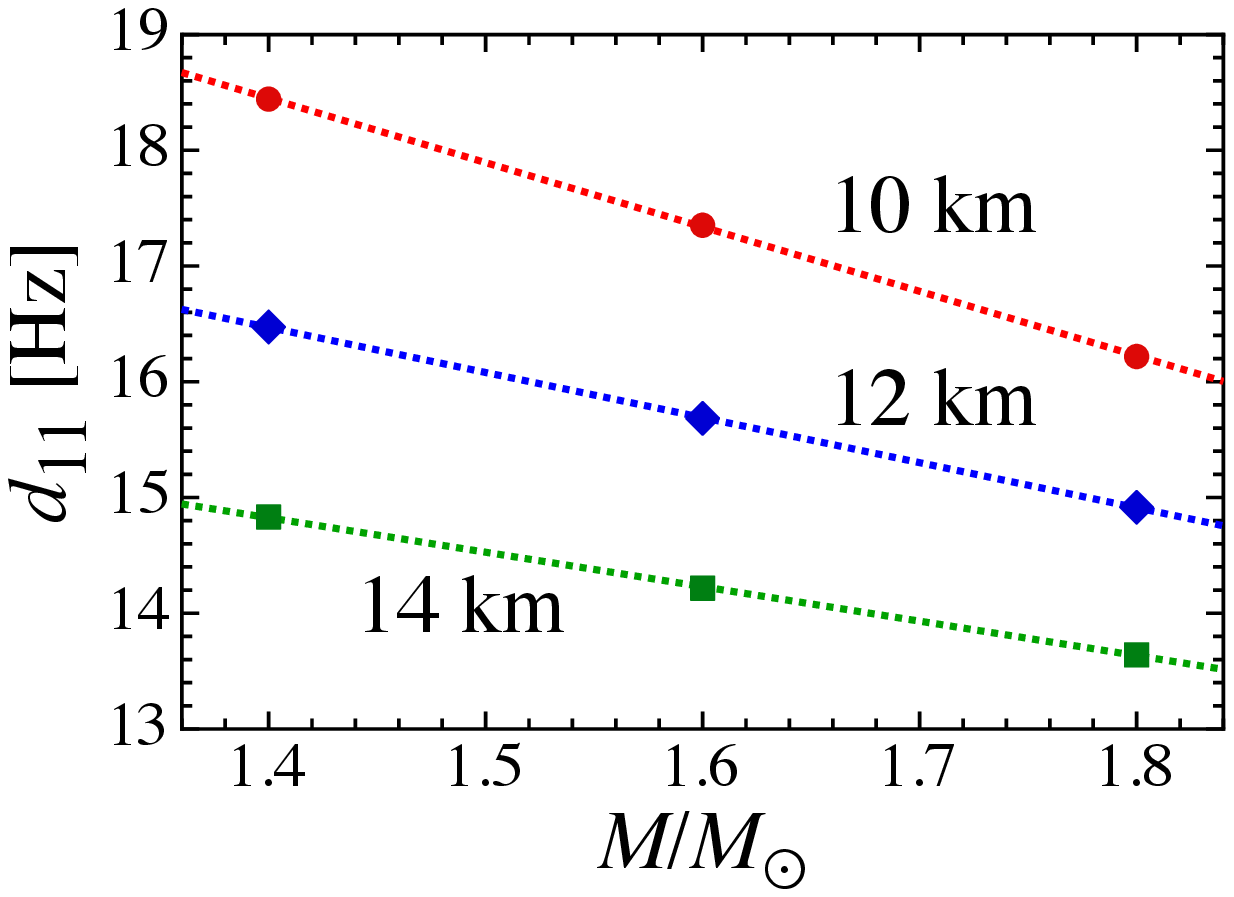} &
\includegraphics[scale=0.42]{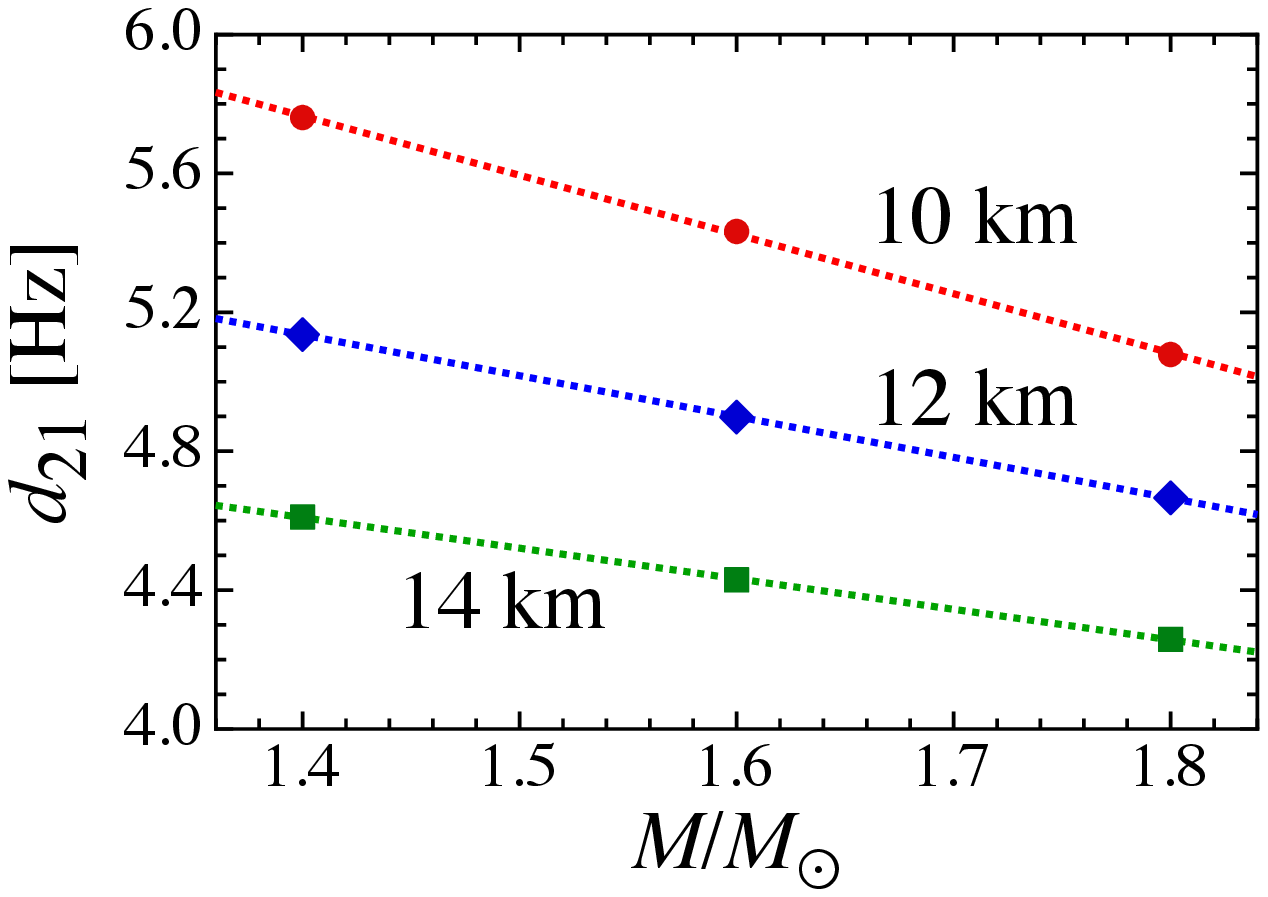} 
\end{tabular}
\end{center}
\caption{
The exact values of $d_{ij}$ obtained from the fitting (\ref{eq:fit1}) are shown for the various stellar models. The circles, diamonds, and squares correspond to the stellar models with $R=10$, 12, and 14 km, respectively, while the dotted lines denote the linear fitting expressed by Eq. (\ref{eq:dij-M}).
}
\label{fig:dij-M}
\end{figure*}

We have successfully expressed ${}_\ell t_0$ as a function of $L$ and $\ell$, where the fitting constants $d_{ij}$ depend on $M$ and $R$. Next, we fit $d_{ij}$ as a function of $M$. In Fig. \ref{fig:dij-M}, we show the exact values of $d_{ij}$ obtained from the fitting (\ref{eq:fit1}) for various stellar models. From this figure, one can observe that the values of $d_{ij}$ would be expressed well as a linear function of $M$ for each stellar radius. Namely, we can write down $d_{ij}$ as
\begin{equation}
  d_{ij} = e_{ij}^0 + e_{ij}^1\left(\frac{M}{M_\odot}\right),  \label{eq:dij-M}
\end{equation}
where $e_{ij}^k$ is a fitting constant for $i=0, 1, 2$, $j=0,1$, and $k=0,1$, depending on $R$. For reference, the obtained linear fitting for $d_{ij}$ is also shown in Fig. \ref{fig:dij-M} by the dotted lines.

\begin{figure*}
\begin{center}
\begin{tabular}{cc}
\includegraphics[scale=0.42]{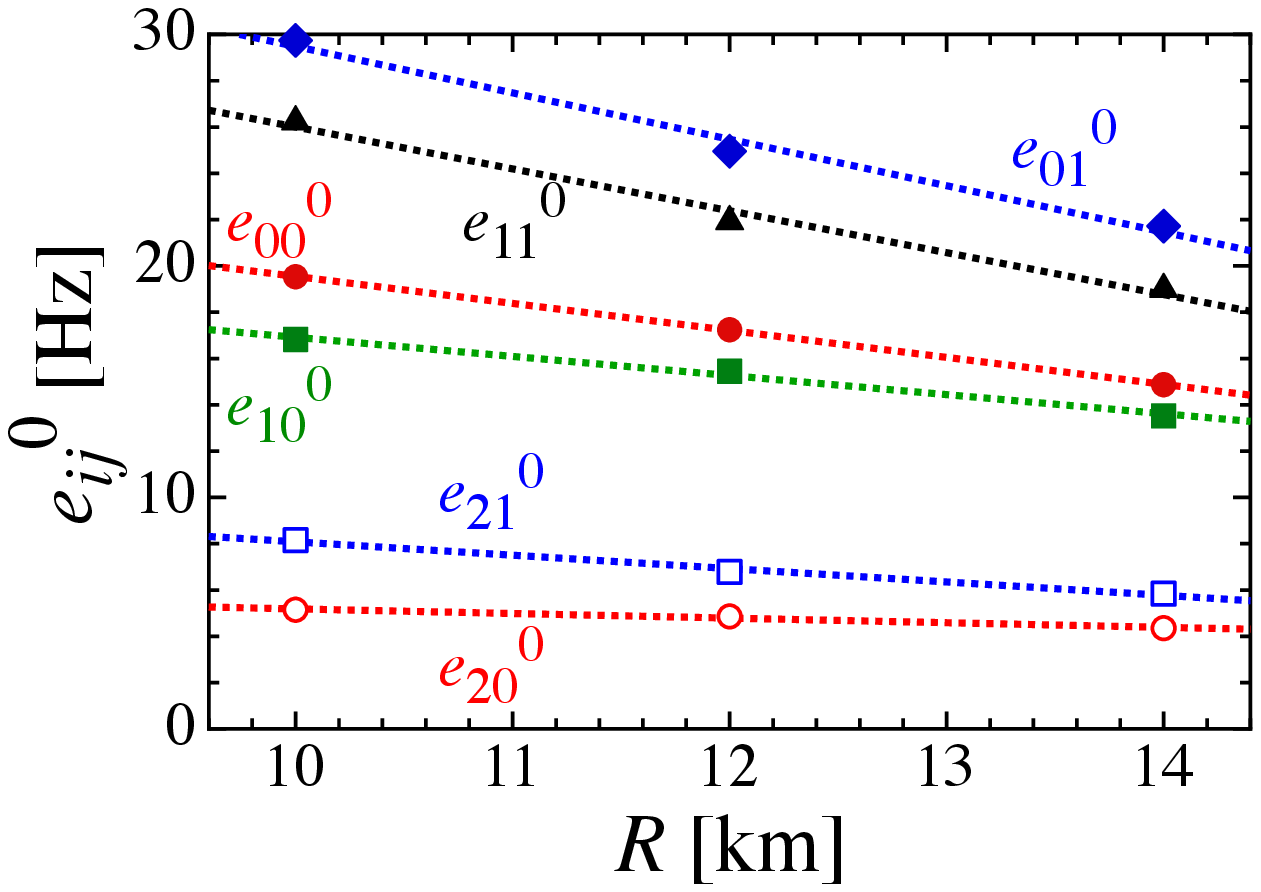} &
\includegraphics[scale=0.42]{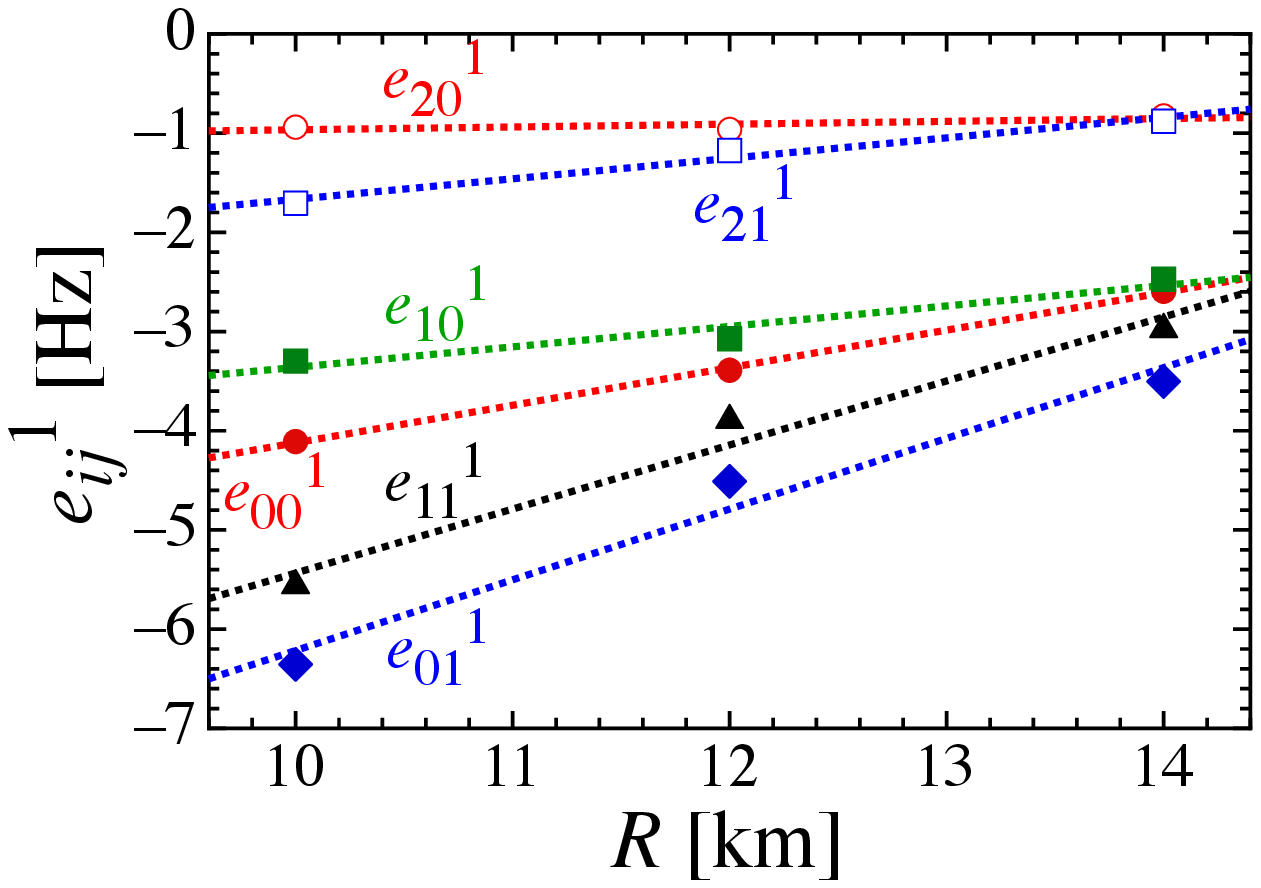} 
\end{tabular}
\end{center}
\caption{
The exact values of $e_{ij}^k$ obtained from the fitting (\ref{eq:dij-M}) are shown as a function of $R$. The filled circles, diamonds, filled squares, triangles, open circles, and open squares correspond to $e_{00}^k$, $e_{01}^k$, $e_{10}^k$, $e_{11}^k$, $e_{20}^k$, and $e_{21}^k$ for $k=0,1$. The dotted lines denote the linear fitting expressed by Eq. (\ref{eq:eijk-R}).
}
\label{fig:eijk-R}
\end{figure*}

Finally, we fit the coefficients $e_{ij}^k$ as a function of $R$. In Fig. \ref{fig:eijk-R}, we show the exact values of $e_{ij}^k$ obtained from the linear fitting (\ref{eq:dij-M}) as a function of $R$. From this figure, it is found that $e_{ij}^k$ is almost a linear function of $R$, such as
\begin{equation}
   e_{ij}^k = f_{ij}^{k0} + f_{ij}^{k1}\left(\frac{R}{10\; {\rm km}}\right),   \label{eq:eijk-R}
\end{equation}
where $f_{ij}^{kl}$ is a fitting constant for $i=0, 1, 2$, $j=0,1$, $k=0,1$, and $l=0,1$. In the same figure, we also plot the linear fitting given by Eq. (\ref{eq:eijk-R}) with the dotted lines. Actually, there is a little deviation of the exact values of $e_{ij}^k$ from the linear fitting, but this deviation could be due to accumulations of small errors in the other fittings for $c_{\ell,i}$ and $d_{ij}$. The obtained fitting coefficients in Eq. (\ref{eq:eijk-R}), $f_{ij}^{kl}$, are shown in Table \ref{tab:f_ijkl}.

\begin{table}[htbp]
\begin{center}
\leavevmode
\caption{Fitting parameters, $f_{ij}^{kl}$, in the empirical formula of crustal torsional oscillations.
}
\begin{tabular}{ cc|c|c|crc  }
\hline\hline
  & $c_{\ell,i}$ & $d_{ij}$ & $e_{ij}^k$ & & $f_{ij}^{kl}$ [Hz]   &  \\
\hline
  & $c_{\ell,0}$ & $d_{00}$  &  $e_{00}^0$  & $f_{00}^{00}=$ & $31.206$ & \\
  &                     &                 &                      & $f_{00}^{01}=$ & $-11.659$ & \\
  &                     &                 &   $e_{00}^1$ & $f_{00}^{10}=$ & $-7.9095$ & \\
  &                     &                 &                      & $f_{00}^{11}=$ &  $3.7875$ & \\
  &                     & $d_{01}$  &  $e_{01}^0$  & $f_{01}^{00}=$ & $49.538$ & \\
  &                     &                 &                      & $f_{01}^{01}=$ & $-20.056$ & \\
  &                     &                 &   $e_{01}^1$ & $f_{01}^{10}=$ & $-13.344$ & \\
  &                     &                 &                      & $f_{01}^{11}=$ & $7.1285$ & \\
\hline
  & $c_{\ell,1}$ & $d_{10}$  &  $e_{10}^0$  & $f_{10}^{00}=$ & $25.173$ & \\
  &                     &                 &                      & $f_{10}^{01}=$ & $-8.2560$ & \\
  &                     &                 &   $e_{10}^1$ & $f_{10}^{10}=$ & $-5.4237$ & \\
  &                     &                 &                      & $f_{10}^{11}=$ & $2.0630$ & \\
  &                     & $d_{11}$  &  $e_{11}^0$  & $f_{11}^{00}=$ & $44.099$ & \\
  &                     &                 &                      & $f_{11}^{01}=$ & $-18.097$ & \\
  &                     &                 &  $e_{11}^1$  & $f_{11}^{10}=$ & $-11.889$ & \\
  &                     &                 &                      & $f_{11}^{11}=$ & $6.4545$ & \\
\hline
  & $c_{\ell,2}$ & $d_{20}$  &  $e_{20}^0$  & $f_{20}^{00}=$ & $7.1755$ & \\
  &                     &                 &                      & $f_{20}^{01}=$ & $-1.9912$ & \\
  &                     &                 &   $e_{20}^1$ & $f_{20}^{10}=$ & $-1.2493$ & \\
  &                     &                 &                      & $f_{20}^{11}=$ & $0.2830$ & \\
  &                     & $d_{21}$  &  $e_{21}^0$  & $f_{21}^{00}=$ & $13.862$ & \\
  &                     &                 &                      & $f_{21}^{01}=$ & $-5.7802$ & \\
  &                     &                 &  $e_{21}^1$  & $f_{21}^{10}=$ & $-3.7306$ & \\
  &                     &                 &                      & $f_{21}^{11}=$ & $2.0638$ & \\
\hline\hline
\end{tabular}
\label{tab:f_ijkl}
\end{center}
\end{table}

\begin{figure*}
\begin{center}
\begin{tabular}{ccc}
\includegraphics[scale=0.42]{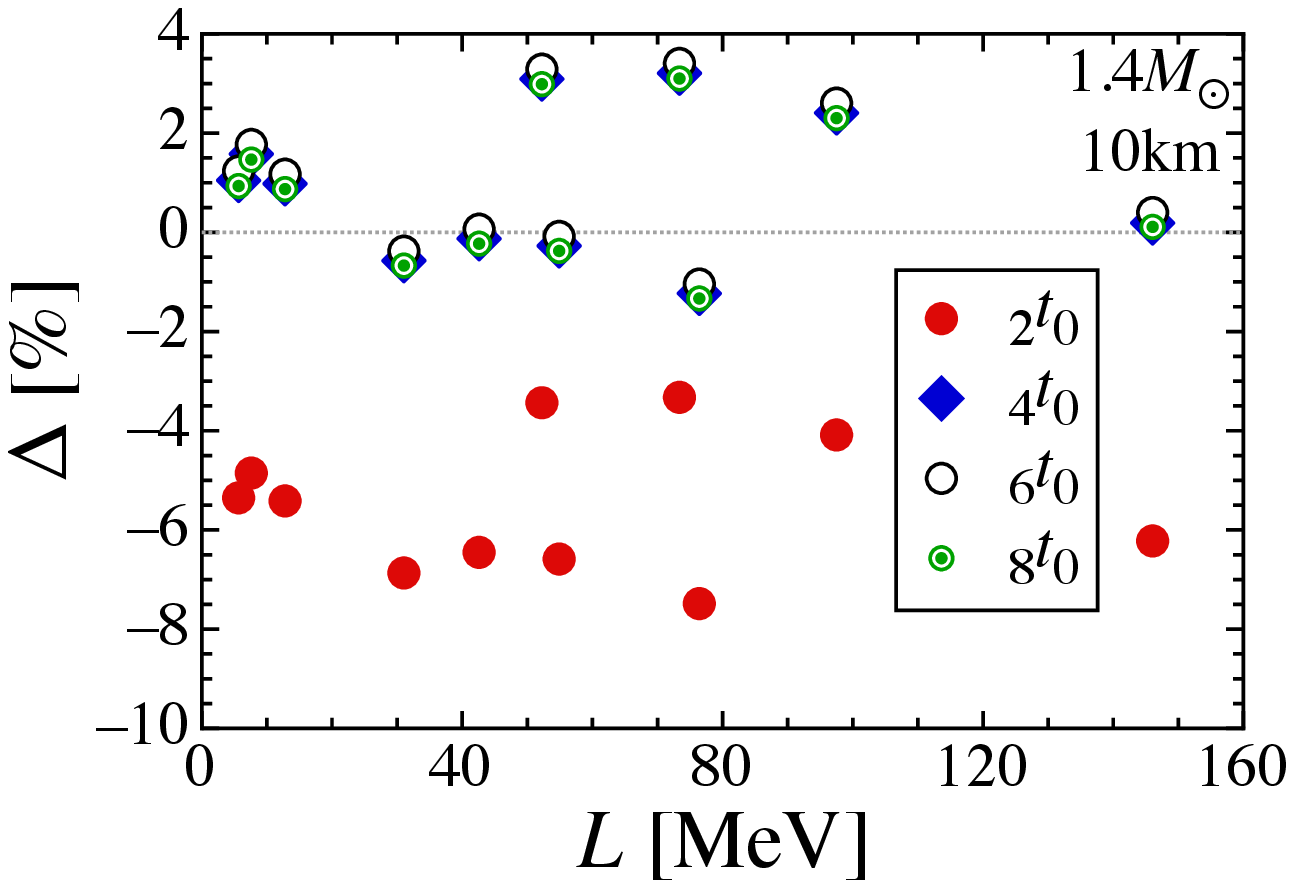} &
\includegraphics[scale=0.42]{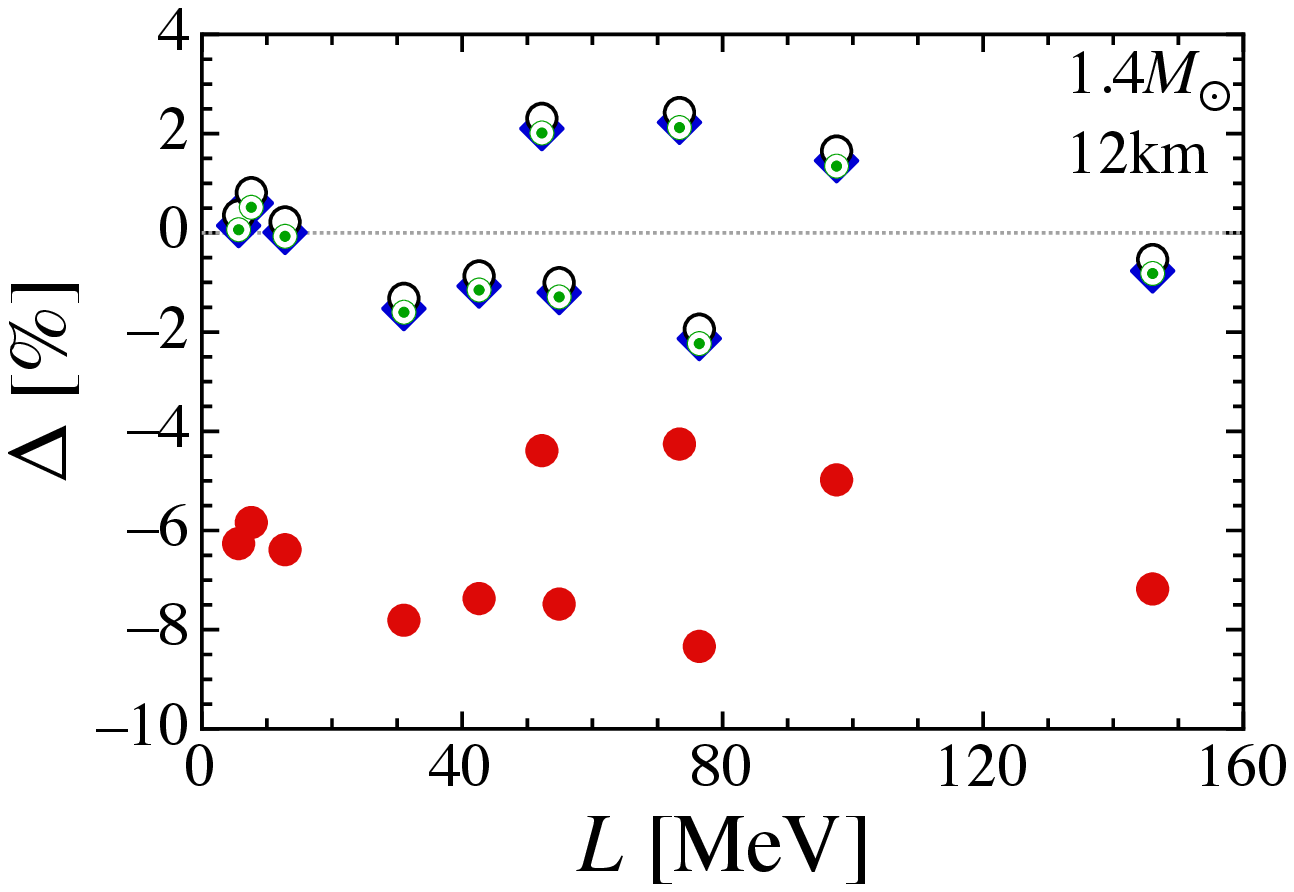} &
\includegraphics[scale=0.42]{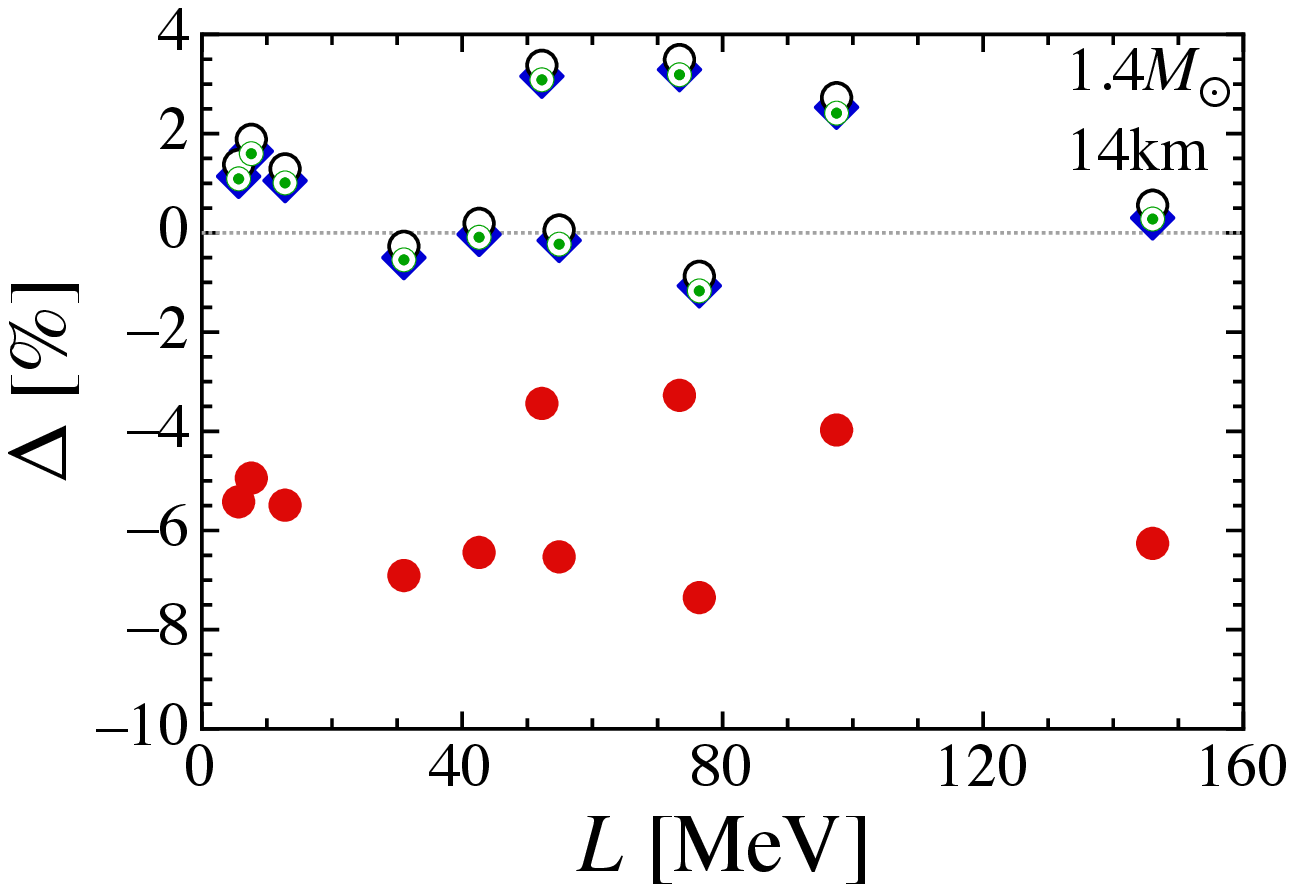} \\ 
\includegraphics[scale=0.42]{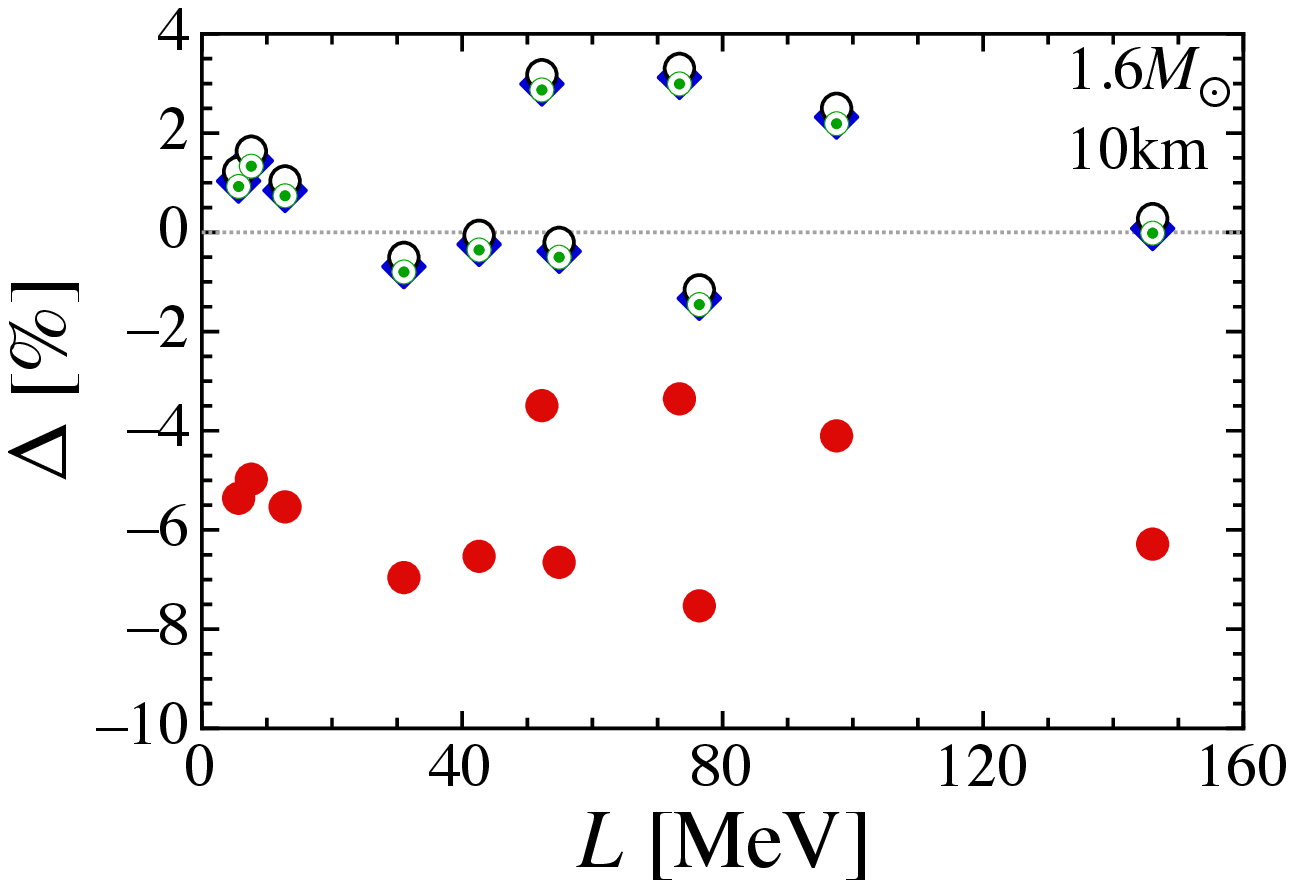} &
\includegraphics[scale=0.42]{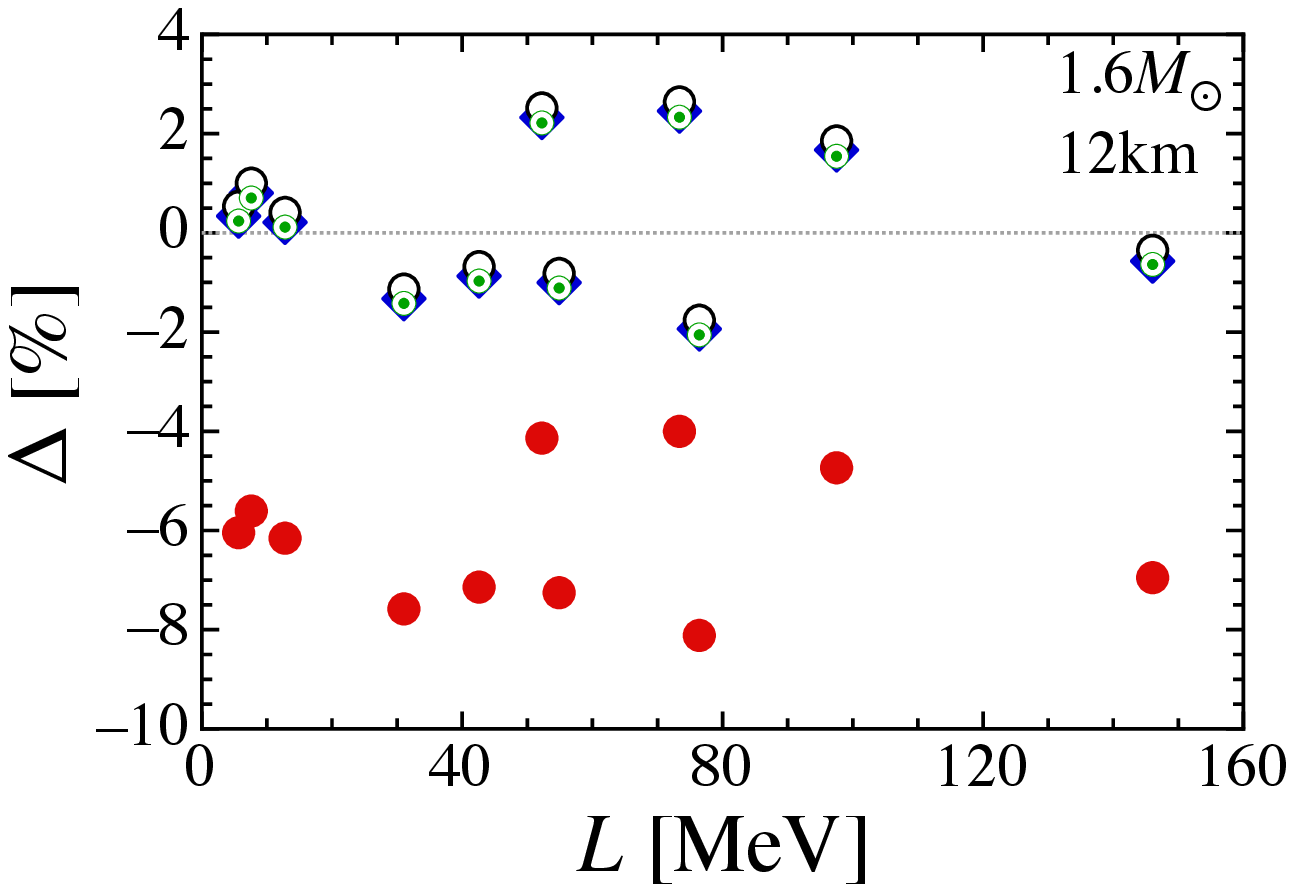} &
\includegraphics[scale=0.42]{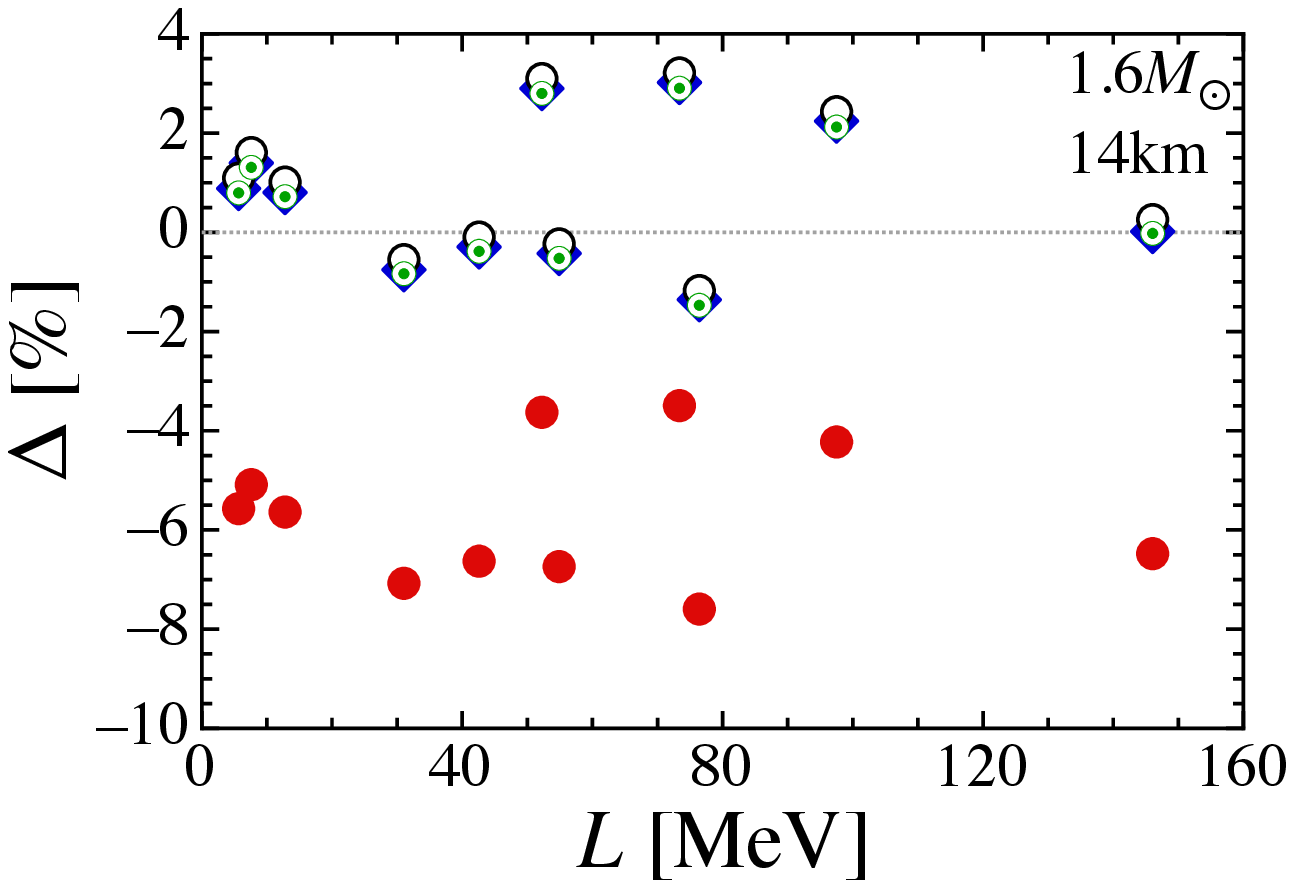} \\
\includegraphics[scale=0.42]{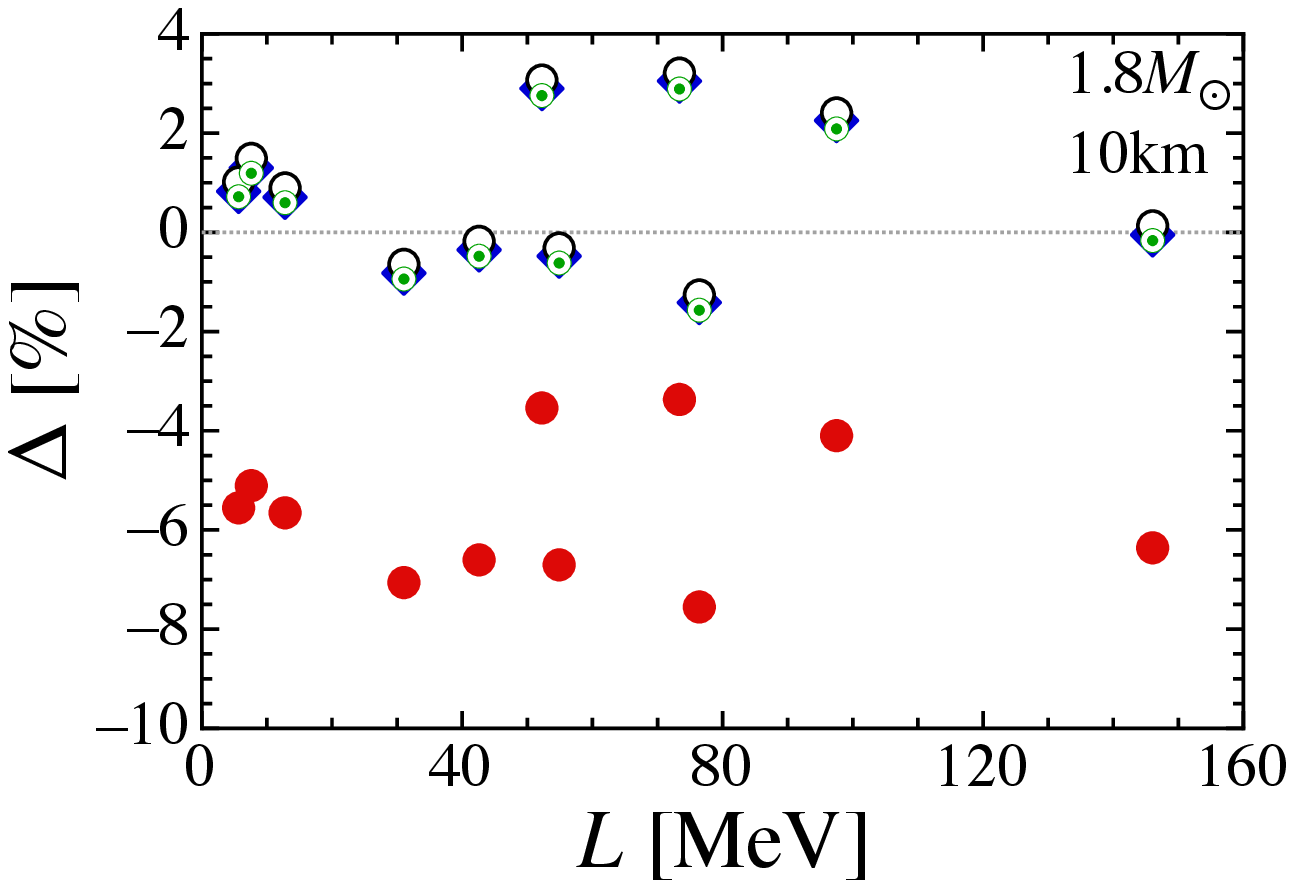} &
\includegraphics[scale=0.42]{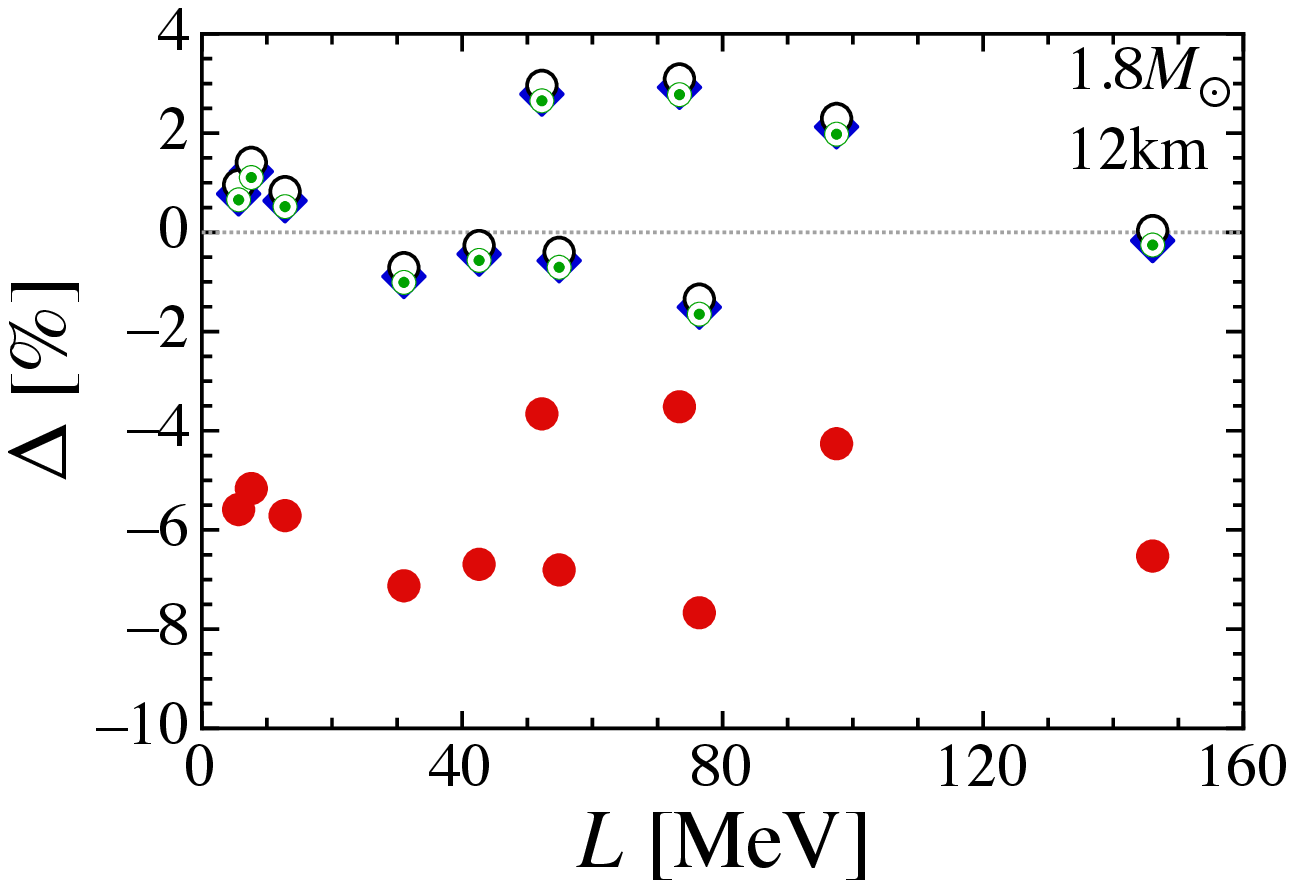} &
\includegraphics[scale=0.42]{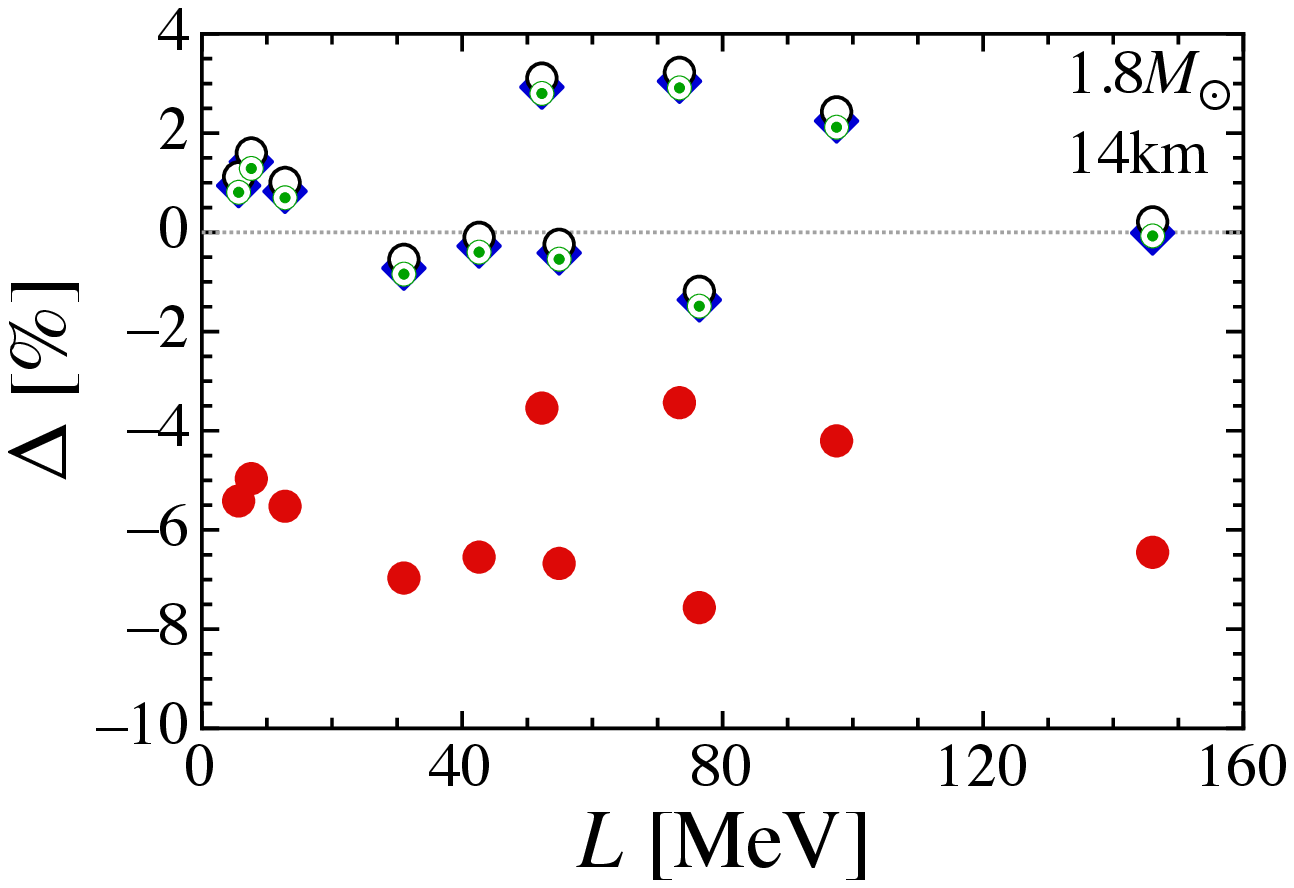} 
\end{tabular}
\end{center}
\caption{
Relative deviations of the frequencies of fundamental torsional oscillations from those estimated with the empirical formula are shown as a function of the slope parameter $L$ for various neutron star models, where the filled circles, diamonds, open circles, and double circles correspond to the frequencies with $\ell=2$, 4, 6, and 8, respectively. 
}
\label{fig:D-MR}
\end{figure*}

At last, we can get the empirical formula of the fundamental frequencies of crustal torsional oscillations as a function of $L$, $\ell$, $M$, and $R$, i.e., ${}_\ell t_0 = {}_\ell t_0(L,\ell,M,R)$, which are given by Eqs. (\ref{eq:lt0}), (\ref{eq:fit1}), (\ref{eq:dij-M}), and (\ref{eq:eijk-R}). Now, we check how accurately this empirical formula can estimate the fundamental frequencies, ${}_\ell t_0$. For this purpose, we calculate the relative deviation, $\Delta$, defined as
\begin{equation}
   \Delta = \frac{{}_\ell t_0^{(n)} - {}_\ell t_0^{(e)}}{{}_\ell t_0^{(n)}},   \label{eq:D-MR}
\end{equation}
where ${}_\ell t_0^{(n)}$ and ${}_\ell t_0^{(e)}$ denote the frequencies of fundamental torsional oscillations calculated numerically and those estimated with the empirical formula, respectively, for various stellar models. The values of $\Delta$ for various stellar models are shown in Fig. \ref{fig:D-MR}, where the filled circles, diamonds, open circles, and double circles denote the fundamental frequencies with $\ell=2$, 4, 6, and 8. In the figure, the top, middle, and bottom panels correspond to the results for the stellar models with $M/M_\odot=1.4$, 1.6, and 1.8, while the left, middle, and right panels correspond to the stellar models with $R=10$, 12, and 14 km. The deviations in the fundamental frequencies with $\ell=2$ are the largest in any of the stellar models, compared with the deviations with different values of $\ell$. Even so, by adopting the empirical formula  one can estimate ${}_2 t_0$ with less than $9\%$ accuracy. Additionally, it should be noted that ${}_\ell t_0$ for $\ell > 2$ can be estimated with less than a few percent, i.e., the values of $\Delta$ for $\ell>2$ become less than $3.5\%$ independently of the stellar models and $L$. The decreased accuracy of the empirical formula for $\ell=2$ may come from the incompatibility in the fitting for the dependence of $\ell$ given by Eq. (\ref{eq:fit1}). In fact, $\Delta$ for $\ell=2$ shown in Fig. \ref{fig:D-MR} is  similar to the values of $\delta_i$ for $\ell=2$ shown in Fig. \ref{fig:delta}. Anyway, we can say that the empirical formula obtained in this paper can work well for a wide range of $L$, $\ell$, $M$, and $R$. Furthermore, we remark that the results in Fig. \ref{fig:D-MR} are derived for various values of $K_0$ in the range of $180\le K_0 \le 360$ MeV (see Table 1 in Ref. \cite{SNIO2013b} for exact values of $K_0$). That is, we can confirm that ${}_\ell t_0$ is almost irrelevant to $K_0$, as shown in the previous study \cite{SNIO2012}. Finally, we remark how the error on the estimation of frequencies affects the determination of the value of $L$ if one could know the stellar mass and radius, which may be determined via additional observations. Using our fitting formula for the stellar model with $M=1.4M_\odot$ and $R=12$ km as a typical neutron star model, one can inversely estimate the value of $L$ with $+6.3\%$ ($-6.1\%$) accuracy at $L=50$ MeV, and with $+3.7\%$ ($-3.4\%$) accuracy at $L=100$ MeV, when one would estimate the frequencies of fundamental crustal oscillations with $-3\%$ ($+3\%$) accuracy.

\section{Conclusion}
\label{sec:IV}

The crustal torsional oscillations directly reflect the crust properties of neutron stars. In practice, by identifying the observed frequencies of neutron stars with the crustal torsional oscillations, one might be able to extract information about the crustal  properties. However, in practice, it is difficult to directly extract information from the observed frequencies, because the frequencies of crustal torsional oscillations generally depend on not only the crust EOS but also the stellar mass and radius. In particular, the core EOS is also strongly associated with the determination of the stellar mass and radius, while there are still many uncertainties in the core EOS depending on the nuclear models and/or the compositions. Thus, in order to purely discuss the crustal torsional oscillations and avoid such uncertainties in the core EOS, we constructed the curst equilibrium models by integrating from the stellar surface up to the crust basis, where we had to prepare two parameters, i.e., the stellar mass and radius. As a result, we successfully derived the empirical formula expressing the fundamental frequencies of crustal torsional oscillations as a function of stellar mass ($M$), radius ($R$), the slope parameter of the nuclear symmetry energy ($L$), and the angular index of the oscillations ($\ell$) for $1.4\le M/M_\odot \le 1.8$, $10\le R\le 14$ km, $0\le L\le 160$ MeV, and $\ell \ge 2$, which are almost independent of the incompressibility in the crust EOS $(K_0)$ in the range of $180 \le K_0\le 360$ MeV. Additionally, we confirmed that by adopting the derived empirical formula one can estimate the frequencies with less than $9\%$ accuracy for $\ell=2$ and less than a few percent accuracy for $\ell>2$. In this paper, we focused only on the fundamental oscillations, but one might be able to extract information about $K_0$ and the crust thickness via the overtones of crustal torsional oscillations \cite{SNIO2012}. In fact, the overtones of crustal torsional oscillations are known to depend on the crust thickness as well as the stellar radius, while the crust thickness might be associated with not only $L$ but also $K_0$, because the both parameters ($L$ and $K_0$) correspond to a kind of stiffness of the EOS. In any case, when discussing the overtones of crustal torsional oscillations one should take into account the effects of the so-called pasta phase between the crust region composed of spherical nuclei and the core region, which will be discussed elsewhere. Furthermore, due to the existence of magnetic fields (which were neglected in this paper) the stellar oscillation might be more complicated, where the crustal oscillations can be coupled with the magnetic oscillations. Even so, the crustal oscillations could be almost independent of such effects, unless the strength of the magnetic field becomes grater than the critical strength, such as $\sim 10^{15}$ G \cite{CK2011,Gabler2011,Gabler2012a}.

\acknowledgments
We are very grateful to K. D. Kokkotas and our referee for giving good suggestions about this work. This work was supported in part by Grant-in-Aid for Young Scientists (B) through Grant No. 26800133 provided by JSPS and by Grants-in-Aid for Scientific Research on Innovative Areas through Grant No. 15H00843 provided by MEXT.



\end{document}